\documentclass[aps,twocolumn,prd]{revtex4}

\usepackage{graphicx}
\usepackage{threeparttable}

\def\be{\begin{equation}}
\def\ee{\end{equation}}
\def\bea{\begin{eqnarray}}
\def\eea{\end{eqnarray}}

\def\cmm2{{\,\rm cm^{-2}}}
\def\cm2{{\,{\rm cm}^2}}
\def\cmm3{{\,{\rm cm}^{-3}}}
\def\gcmm3{{\,{\rm g\,cm^{-3}}}}

\def\fun#1#2{\lower3.6pt\vbox{\baselineskip0pt\lineskip.9pt
  \ialign{$\mathsurround=0pt#1\hfil##\hfil$\crcr#2\crcr\sim\crcr}}}

\def\p3m{P$^3$M}

\def\ga{\mathrel{\mathpalette\fun >}}
\def\fun#1#2{\lower3.6pt\vbox{\baselineskip0pt\lineskip.9pt
  \ialign{$\mathsurround=0pt#1\hfil##\hfil$\crcr#2\crcr\sim\crcr}}}

\def\neff{$N_{\rm eff}$}

\def\yp{Y_{\rm P}}

\begin{document}

\def\affilmrk#1{$^{#1}$}
\def\affilmk#1#2{$^{#1}$#2;}

\def\ucd{1}
\def\uc{2}
\def\ucb{3}

\bibliographystyle{apsrev}
\title{How Massless Neutrinos Affect the Cosmic Microwave
  Background Damping Tail}
\author{
Zhen\ Hou\affilmrk{\ucd},
Ryan\ Keisler\affilmrk{\uc}, 
Lloyd\ Knox\affilmrk{\ucd},
Marius\ Millea\affilmrk{\ucd} and
Christian\ Reichardt\affilmrk{\ucb}
}
\affiliation{
\parshape 1 -3cm 24cm
\affilmk{\ucd}{Department of Physics, One Shields Avenue,
University of California, Davis, California 95616, USA}
\affilmk{\uc}{Department of Astronomy \& Astrophysics, 
University of Chicago, 5640 S. Ellis Avenue, Chicago, Illinois 60637, USA}
\affilmk{\ucb}{Department of Physics,
University of California, Berkeley, CA 94720, USA}
}
\date{\today}

\begin{abstract}
We explore the physical origin and robustness of constraints on the energy
density in relativistic species prior to and during recombination,
often expressed as constraints on an effective number of neutrino
species, $N_{\rm eff}$.  Constraints from current data
combination of {\em Wilkinson Microwave Anisotropy Probe} ({\em WMAP}) and
South Pole Telescope (SPT) are almost entirely due to the impact of the
neutrinos on the expansion rate, and how those changes to the expansion rate
alter the ratio of the photon diffusion scale to the sound horizon scale at
recombination.  We demonstrate that very little of the constraining power
comes from the early Integrated Sachs-Wolfe (ISW) effect, and also provide a
first determination of the amplitude of the early ISW effect.  Varying
the fraction of baryonic mass in Helium, $\yp$, also changes the ratio
of damping to sound-horizon scales.  We discuss the physical effects
that prevent the resulting near-degeneracy between $N_{\rm eff}$ and
$\yp$ from being a complete one.  Examining light element abundance
measurements, we see no significant evidence for evolution of $N_{\rm eff}$ and
the baryon-to-photon ratio from the epoch of big bang nucleosynthesis
to decoupling.  Finally, we consider measurements of the
distance-redshift relation at low to intermediate redshifts 
and their implications for the value of $N_{\rm eff}$.  
\end{abstract}
  \maketitle

 \section{Introduction}

High-resolution observations of the cosmic microwave background (CMB)
temperature anisotropy are providing a precise measurement
of the damping tail of CMB power spectrum, shedding light on the physical
conditions during recombination and back into the radiation dominated era.  The
measurements have
revealed somewhat less fluctuation power at small angular scales than expected
in the standard cosmological model \citep{reichardt09, das11, dunkley10,
keisler11}. Many papers \citep{dunkley10, deHolanda10, fischler10, krauss10,
galli10, nakayama10, hamann10, keisler11} have considered the possibility that
this deficit of power is due to extra (dark) relativistic species
\footnote{\citet{jungman96} first pointed
out that the number of species of relativistic neutrinos could be measured from
CMB.}, such as nearly massless sterile neutrinos.  Constraints are usually
expressed in terms of an effective number of neutrinos, $N_{\rm eff}$
\footnote{The effective number of neutrinos is defined so that the neutrino and
photon energy densities are related by $\rho_\nu = N_{\rm eff} \,7/8\,
(4/11)^{4/3} \rho_\gamma$.}.

Note that these constraints apply to any weakly interacting or
non-interacting species that is relativistic at recombination.
Hypothesized additional species include sterile neutrinos, sub-eV mass
axions as in, e.g., \citet{hannestad10} and those arising in many other
extensions of the standard model such as in \citet{fischler11}.

Interest in the number of light degrees of freedom is further
stimulated by i) recent inferences of the primordial Helium abundance
which are larger, and with larger uncertainties than previous analyses
\cite{izotov10,aver10,aver11}, ii) evidence for additional (sterile)
neutrino species from laboratory-produced neutrinos \cite{miniboone10}
and reactor-produced neutrinos \cite{mention11} and iii) a slight tension between determinations of
distance vs. redshift at very low redshifts (essentially measurements
of $H_0$ \cite{riess11}) and those at low to intermediate redshifts
that use the baryon acoustic oscillation (BAO) feature in the galaxy power
spectrum as a CMB-calibrated standard ruler \cite{mehta12,anderson12}.

Given this interest, and impending improvements to the damping tail
measurements from the SPT \citep{keisler11} (K11 hereafter) and the {\it Planck}
satellite, a pedagogical exposition of the impact of $N_{\rm eff}$ on the CMB is
quite timely.  The relevant physics is beautifully simple and deserves
to be as well (and as broadly) understood as the constraints on
$N_{\rm eff}$ from big bang nucleosynthesis (BBN) considerations.  Focusing on
the physics behind the constraints also allows one to
understand their more general implications.  The only important
assumptions about the relativistic degrees of freedom that go into the
CMB predictions is that their interactions are negligible and they are
massless.

Despite the analyses provided by
\citet{huwhite96} and \citet{bashinsky04} we find confusion in the literature about {\em
why} increasing $N_{\rm eff}$ reduces small-scale anisotropy.  The sensitivity
to \neff\ from high-resolution CMB observations is due to the impact of the mean
relativistic energy density on the Universe's expansion rate prior to (and
during) the epoch of photon-baryon
decoupling.  As we demonstrate below, neutrino perturbations do not play a significant
role \footnote{An exception is the breaking of the $Y_{\rm P}-N_{\rm
    eff}$ degeneracy, to be discussed later.}, nor
do anisotropies induced after decoupling.  To demonstrate the
insignificance of the early Integrated Sachs-Wolfe effect, we
artificially introduce a scaling of its physical value by a parameter
$A_{\rm eISW}$ and examine the constraints on $N_{\rm eff}$ that follow
from marginalizing over $A_{\rm eISW}$.

The constraints on $N_{\rm eff}$ are model dependent; there are other
ways to extend the standard cosmological model to suppress small-scale
power in the cosmic microwave background.  We discuss, in particular,
departures from a power law for the primordial perturbation spectrum,
and allowing the fraction of baryonic mass in Helium, $Y_{\rm P}$, to
be a free parameter.

We also consider the sensitivity of inferences of $N_{\rm eff}$ from
CMB data to assumptions about BBN and measurements of light-element
abundances.  Allowing $\yp$ to vary freely introduces a degeneracy
that completely changes the mechanism by which $N_{\rm eff}$ is
constrained by CMB data.  We comment on the origin of constraints in
the $N_{\rm eff}$ - $\yp$ plane.

In Section II we review the analytic explanation for the
origin of constraints on $N_{\rm eff}$ from CMB observations and
demonstrate its quantitative effectiveness in understanding
constraints from current data.  We also examine the model
dependence of constraints on $N_{\rm eff}$.  In Section III we
consider the impact of the early ISW effect on $N_{\rm eff}$
constraints and we discuss our results and conclude in Section IV.

\section{How CMB Observations Constrain Component Densities}

Let us review the sensitivity of the CMB power
spectrum to the densities today of baryons, $\rho_b$, cold dark matter
plus baryons, $\rho_m$, and the dark energy, $\rho_\Lambda$, all
within the context of the $\Lambda$CDM model \footnote{Energy
  densities today are often expressed as, e.g., $ \Omega_b h^2 =
  \rho_b/(1.879 \times 10^{-29}$g/cm$^3$)}. For more details than we
give here, see \citep{hudodelson02,hufukugita01}. In the course of our
review, we will identify directions in parameter space that cause
large changes in probability density.  When we do study variations in
$N_{\rm eff}$ we will do so along orthogonal directions.

\subsection{Sensitivity to the Non-neutrino Components}
The dependence of the CMB on $\rho_b$ arises from the dependence of
the equation-of-state of the pre-recombination plasma on the fraction
of its energy density that comes from baryons.  Increasing the number
of baryons per photon decreases the plasma's pressure to density ratio
($P/\rho$).  The resulting shift in the equilibrium point between
gravitational and pressure forces alters the ratio of even peak
heights to odd peak heights.  We know the energy density of photons
very well from measuring the CMB photon spectrum \citep{mather90}, and
can thus infer $\rho_b$ from the ratio of even to odd peak heights.

The dependence on $\rho_m$ arises from the sensitivity of the
evolution of a Fourier mode amplitude to the fraction of the energy
density contributed by non-relativistic matter when the mode's
wavelength is equal to the Hubble radius.  This ratio depends on
$\lambda/r_{EQ}$ where $\lambda$ is the comoving wavelength,
$r_{\rm EQ} = H^{-1}_{EQ}/a_{\rm EQ}$ defines the comoving Hubble
radius at the time when the matter density equals the radiation
density (an epoch denoted by `EQ') and $a$ is the scale factor parameterizing the
expansion of the Universe.
The amplitude of a mode projecting to angular scale $\theta$
depends on $\theta/\theta_{\rm EQ}$ where $\theta_{\rm EQ} = r_{\rm
 EQ}/D_A$ because $\theta/\theta_{\rm EQ} = \lambda/r_{\rm EQ}$.
Since the amplitude is a strong function of this ratio, there
is strong sensitivity to $\theta_{\rm EQ}$ and therefore to $z_{\rm EQ}$ ($1+z = 1/a$)
since, assuming the dark energy is a cosmological constant,
$\theta_{\rm EQ} = I(\Omega_m) /\sqrt{1+z_{\rm EQ}}$ where $I(\Omega_m)$ is a very
slowly-varying function of $\Omega_m$\footnote{$\Omega_m \equiv
 8 \pi G\rho_m/(3 H_0^2)$.}.

Thus the CMB power spectrum is sensitive to $1+ z_{\rm EQ} =
\rho_m/\rho_r$.  If we {\em assume} the standard radiation content ($N_{\rm
  eff} = 3.046$, $\rho_r = \rho_\gamma + \rho_\nu$), then a constraint
on $1+z_{\rm EQ}$ directly constrains $\rho_m$.  However, if we are
allowing $N_{\rm eff}$ to vary, we should study its effects at fixed
$1+z_{\rm EQ}$ since this quantity is well
constrained by the data.  Prior to the CMB damping scale
measurements, there were already hints of high $N_{\rm eff}$ by
combining the $1+z_{\rm EQ}$ constraint from {\it WMAP} with
late-time observables sensitive to the matter density \cite{komatsu09}.  

We now consider $\rho_\Lambda$.  The angular
scales of the acoustic peaks are highly sensitive to the angular size
of the sound horizon, $\theta_s = r_s/D_A$.  Thus,
$\theta_s$ is very precisely determined by the data.  Given
$r_s$, we could infer $D_A$.  In a universe with zero mean spatial
curvature, $D_A = \int c \, dt/a$ from the time of last scattering to today,
and depends only on $\rho_m$ and $\rho_\Lambda$.  We do know $r_s$, to some
degree, from our determination of $\rho_b$ and $\rho_m$ as described above.
These densities determine the history of the sound speed, $c_s$\footnote{$c^2_s
= \partial P/\partial \rho = 1/[3(1+R)]$ with $R
= 3 \rho_b(z)/(4\rho_\gamma(z))$} and the expansion rate, allowing us to
calculate $r_s$ since it depends on no other parameters in the six-parameter
$\Lambda$CDM model.  This determination of $r_s$ allows for a constraint on
$\rho_\Lambda$.

\subsection{The Effect of Relativistic and Dark Degrees of
  Freedom}

\begin{figure}
	\begin{center}
	\includegraphics[width=0.48\textwidth, trim=2.5cm 13.2cm 10.3cm
        2cm]{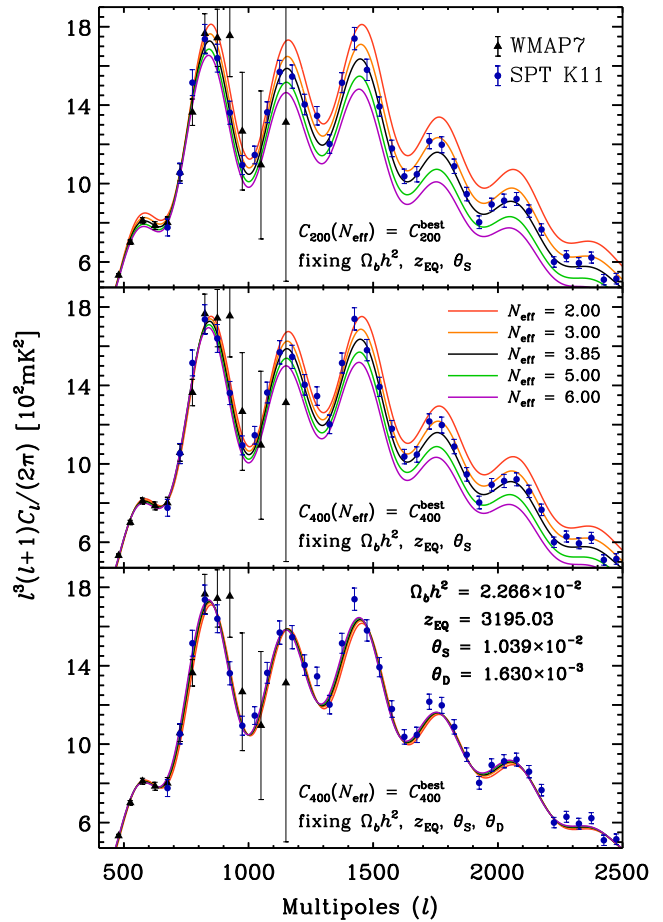}
	\end{center}
	\caption{{\em Top panel:} {\it WMAP} and SPT power spectrum
          measurements, and theoretical power spectra normalized at
          $\ell = 200$.  The black (central) curve is for the best-fit
          $\Lambda$CDM + $N_{\rm eff}$ model assuming BBN consistency.  The
          other
          model curves are for $N_{\rm eff}$ varying
          from 2 to 6 with $\rho_b$, $\theta_s$, and $z_{\rm EQ}$ held
          fixed.  Larger $N_{\rm eff}$ corresponds to lower power.
          {\em Central panel:}  Same as above except normalized at
          $\ell = 400$ where the ISW contribution is negligible.  We see most
          of the variation remains.
          {\em Bottom panel:} The same as the central panel except we vary $\yp$
          to keep $\theta_d$ fixed.  The lack of scatter in these
          spectra compared to those in the middle panel demonstrates that
          the effect of $N_{\rm eff}$ on small-scale data is largely captured by its
          impact on the damping scale.  We can also begin to see more
          subtle effects of the neutrinos, most noticeably a phase
          shift in the acoustic oscillations \cite{bashinsky04}. }
	\label{fig:cls}
\end{figure}

Changing $N_{\rm eff}$ only slightly alters the above story about the origins
of the parameter constraints.  We expect inferences of $\rho_b$, $1+z_{\rm
  EQ}$ and $\theta_s$ to be nearly unaffected.  In the top panel of
Fig.~\ref{fig:cls} we thus hold these parameters fixed while varying
$N_{\rm eff}$ from 2 to 6.  We hold $z_{\rm EQ}$ fixed by
increasing the density of cold dark matter as we
increase $N_{\rm eff}$.  We keep $\theta_s$ fixed by changing
$\rho_\Lambda$ to adjust $D_A$.

As can be seen in Fig.~\ref{fig:cls}, increasing $N_{\rm eff}$ along
the chosen direction in parameter space, makes very
little difference at low $\ell$ (exactly as intended) and an increasing
difference at higher $\ell$.  A similar exercise was performed in
\cite{bowen02}, who identified the same parameters to be held fixed, but then
ascribed the relative drop in power toward small scales as due to a
post-decoupling effect known as the ``early integrated Sachs-Wolfe
effect.''  This explanation was summarized and cited in
\cite{galli10}.  We claim, in contrast, that these differences at high
$\ell$ are almost entirely due to increased Silk damping, caused by the
increased expansion rate.  To demonstrate that the
variation is not predominantly due to ISW, in the central panel we have
normalized the spectra at $\ell = 400$ where the ISW effect is
negligible; we see that the high-$\ell$ variation is only
slightly reduced.

Temperature anisotropies on scales smaller than the photon diffusion
length are damped by the diffusion, a phenomenon known as Silk
damping.  Diffusion causes the drop in power toward high $\ell$ and
makes the power spectrum sensitive to the angular scale of the
diffusion length, $\theta_d$.  To second order in $\lambda_{\rm
  mfp}/\lambda$, where $\lambda_{\rm mfp}$ is the photon mean free
path, the temperature fluctuations are suppressed by
$\exp[-(2 r_d/\lambda)^2]$ where the mean squared diffusion
distance at recombination is
\be
\label{eqn:rd}
r_d^2 = \pi^2\int_0^{a_*} \frac{da}{a^3\sigma_T n_e
  H}\left[\frac{R^2 + \frac{16}{15}\left(1+R\right)}{6(1+R^2)}\right]
\ee
where $n_e$ is the number density of free electrons, $\sigma_T$ is the Thompson cross-section, $a_*$ is the
scale factor at recombination (defined below) and the factor in 
square brackets is due to the directional and
polarization dependence of Thompson scattering \citep{kaiser83,zaldarriaga95}.  
Although Eq.~\ref{eqn:rd} is only an approximation 
to the diffusion length, it allows an analytic understanding
of the dependence of this diffusion length on model parameters
\citep{huwhite96}.  

If we approximate $a_*$ as independent of $H$, then $r_d \propto
H^{-0.5}$.  This is as expected for a random walk process: the
distance increases as the square root of time.  Increasing $H$ (which happens
when we increase $N_{\rm eff}$) leads to smaller $r_d$ which would decrease the
amount of damping.  Why do we see, in Fig.~\ref{fig:cls}, the damping
increase as $N_{\rm eff}$ increases?

The answer has to do with how $r_s$ and $D_A$ change to keep $\theta_s$ fixed
despite the increased expansion rate.  The comoving sound horizon is given by
\be\label{eqn:rs}
r_s = \int_0^{t_*} c_s \, dt/a =\int_0^{a_*} \frac{c_s\, da}{a^2 H}.
\ee
Since $r_s \propto 1/H$, it responds even more rapidly to changes in
$H$ than is the case for $r_d$.  To keep $\theta_s$ fixed at the observed value,
$D_A$ must also scale as $1/H$. Since $D_A$ decreases by more than would be
necessary to keep $\theta_d$ fixed, $\theta_d$ increases which means
the damping is increased.  

To look at it another way, if we knew $D_A$
perfectly, we could use $r_s$ to determine $H$ prior to
recombination.  But we do not know $D_A$, largely because we do not
know the value of the cosmological constant, or more generally the
density of the dark energy as a function of the scale factor.
Instead, we can use the two scales together to form a ratio that is
sensitive to $H$, with no dependence on $D_A$:  $\theta_d/\theta_s =
r_d/r_s \propto H^{0.5}$.  

Does this explanation hold together quantitatively?  To demonstrate that what we are seeing
in the power spectrum actually is increased Silk damping (at fixed
$\theta_s$) we experiment with also fixing $\theta_d$ as $N_{\rm eff}$
increases.  The bottom panel of Fig.~\ref{fig:cls} shows how the
angular power spectrum responds to the same variations in $N_{\rm
  eff}$, only now taken at constant $\theta_d$ as well.  When we
remove the $\theta_d$ variation, the impact of the $N_{\rm eff}$
variation almost entirely disappears.  We conclude that the variations
we are seeing in the top panel are indeed due to the impact of $N_{\rm
  eff}$ on the amount of Silk damping.  A very similar demonstration
was provided by \cite{bashinsky04}.

To keep $\theta_d$ fixed as $N_{\rm eff}$ varies, we varied a
parameter whose sole impact is on the number density of electrons: the
primordial fraction of baryonic mass in Helium, $\yp$.  Even as early
as times when 99\% of the photons have yet to last scatter, Helium,
with its greater binding energy than Hydrogen, is almost entirely
neutral.  Thus $n_e = X_e (n_p+n_H) = X_e n_b (1-\yp)$ where the first
equality defines $X_e$ and we have kept $n_b$ (and thus $\rho_b$)
fixed.  The limit of integration in the above equations for $r_s$ and
$r_d$ is only slightly affected by changing $\yp$ and thus $r_s$ is largely unaffected.  
However, the damping length scales with $\yp$ as $r_d \propto (1-\yp)^{-0.5}$.

From our analysis one finds that $r_d/r_s \propto
(1+f_\nu)^{0.25}/\sqrt{1-Y_P}$ where $f_\nu \equiv \rho_\nu/\rho_\gamma$ is
proportional to $N_{\rm eff}$.  The first factor arises because
increasing $H$ at fixed $z_{\rm EQ}$ means $H^2 \propto (1+f_\nu)$.  
Thus as $N_{\rm eff}$ is varied, we know how to change $\yp$ to keep
$r_d/r_s$ (and hence $\theta_d/\theta_s$) fixed.  

However, the above analysis requires a small correction for two reasons.
First, increased expansion, even if we keep $n_e(a)$ fixed, decreases
$a_*$ because we define $a_*$, following \cite{husugiyama95}, such that the
optical depth to Thomson scattering from here to $a_*$ is unity.  Second,
recombination is not a process that occurs in chemical equilibrium.
As emphasized in \cite{zahn03}, increasing the expansion rate leads to
an increase in $n_e(a)$.  By numerically studying these effects, which partially
cancel each other, we find that $r_d/r_s \propto (1+f_\nu)^{m}/\sqrt{1-\yp}$
with $m = 0.28$ rather than 0.25.

Note that when varying $N_{\rm eff}$ in Fig.~\ref{fig:cls}, we also
vary $\yp$ as is expected for standard assumptions about
BBN, as will be explained below.
Following BBN consistency (as opposed to keeping $\yp$ fixed) increases the damping effect by about
30\%.

\begin{figure}
	\centering
	\includegraphics[width=0.48\textwidth, trim=2cm 13cm 10cm
6.9cm]{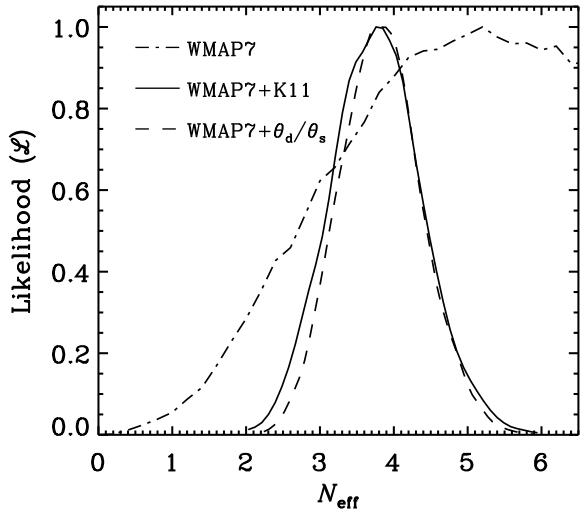}
	\caption{Probability distribution of $N_{\rm
            eff}$ using just {\it WMAP} data, {\it WMAP} + SPT, and
          {\it WMAP} + constraints on $\theta_d/\theta_s$ from {\it
            WMAP} + SPT (see Tab.~\ref{tab:neff_bbn}).
	\label{fig:itsthetadthetas}}
\end{figure}

We should mention that neutrino perturbations do alter the
amplitude of the power spectrum at $l \ga 200$ by a (nearly) constant factor 
\cite{huscott95,bashinsky04}.  The breaking of the
$N_{\rm eff}$, $\rho_m$ degeneracy in {\it WMAP} data at low $N_{\rm
eff}$ is due to the impact of neutrino perturbations, and this is
the effect that allowed for an indirect detection of these
perturbations as reported in \cite{trotta05,debernardis08}.

However, with the inclusion of small-scale data, the perturbations
have lost their significance.  In Fig.~\ref{fig:itsthetadthetas} we
demonstrate that the $N_{\rm eff}$ constraint from {\it WMAP} + SPT 
is well approximated by combining the WMAP7 data with the information
on $\theta_d/\theta_s$ from {\it WMAP} +SPT.  

\begin{figure}
    \centering
    \includegraphics[width=0.48\textwidth, trim=0.5cm 7.5cm 3cm
5.7cm]{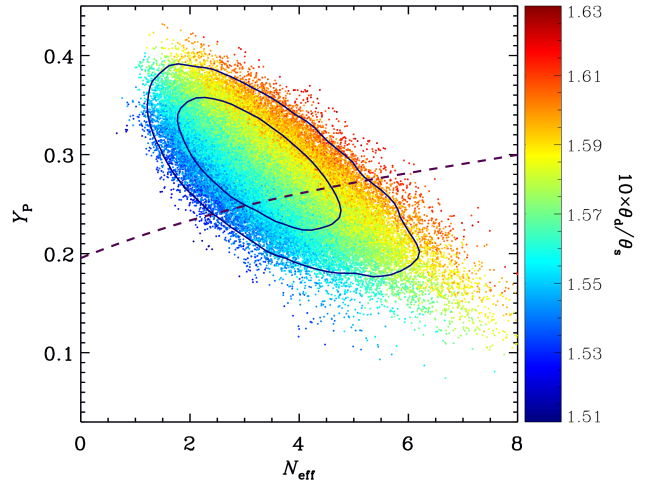}
    \caption{The joint likelihood of $N_{\rm eff}$ and $\yp$ with 68\%
and 95\% confidence contours. Each
scattered point in the figure shows one element of the Markov chain with
color coding the corresponding value of $\theta_d/\theta_s$.  The purple dashed
line is the BBN consistency line.}
    \label{fig:neff_yp}
\end{figure}

Fig.~\ref{fig:neff_yp} provides another way of seeing the importance
of $\theta_d/\theta_s$ to the $N_{\rm eff}$ constraint.  From the
color coding one can see that lines of constant $\theta_d/\theta_s$
run along the major axis of the probability contours.  Further, one
can see that the BBN consistency line cuts nearly perpendicularly
across these lines.  This feature explains why the errors on $N_{\rm eff}$ are
about 30\% smaller if one assumes BBN consistency rather than fixed $\yp$.  If
we abandon BBN consistency and allow $\yp$ to vary freely, then $N_{\rm
eff}$ is allowed to vary along the major axis of the probability contours and
the constraint on $N_{\rm eff}$ loosens considerably, as described in more
detail below.

\subsection{Constraining $N_{\rm eff}$ with $\yp$ free}
\label{subsec:constrain_neff_yp}

We do not have a complete analytic understanding of the closing of the
contours on the major axis (as opposed to the minor axis) in
Fig.~\ref{fig:neff_yp}.  We can turn though to the lowest panel of
Fig.~1 to see that at fixed $\theta_d/\theta_s$ there is indeed some
remaining variation to the power spectra as $N_{\rm eff}$ varies.
At least some of this variation is due to the difference in acoustic
oscillation phase shift that one gets for neutrinos, relative to the
same energy density in photons, due to their free streaming \cite{bashinsky04}. 


\begin{figure}
    \begin{center}
    \includegraphics[width=0.48\textwidth, trim=2.2cm 13.0cm 5.8cm
        2.8cm]{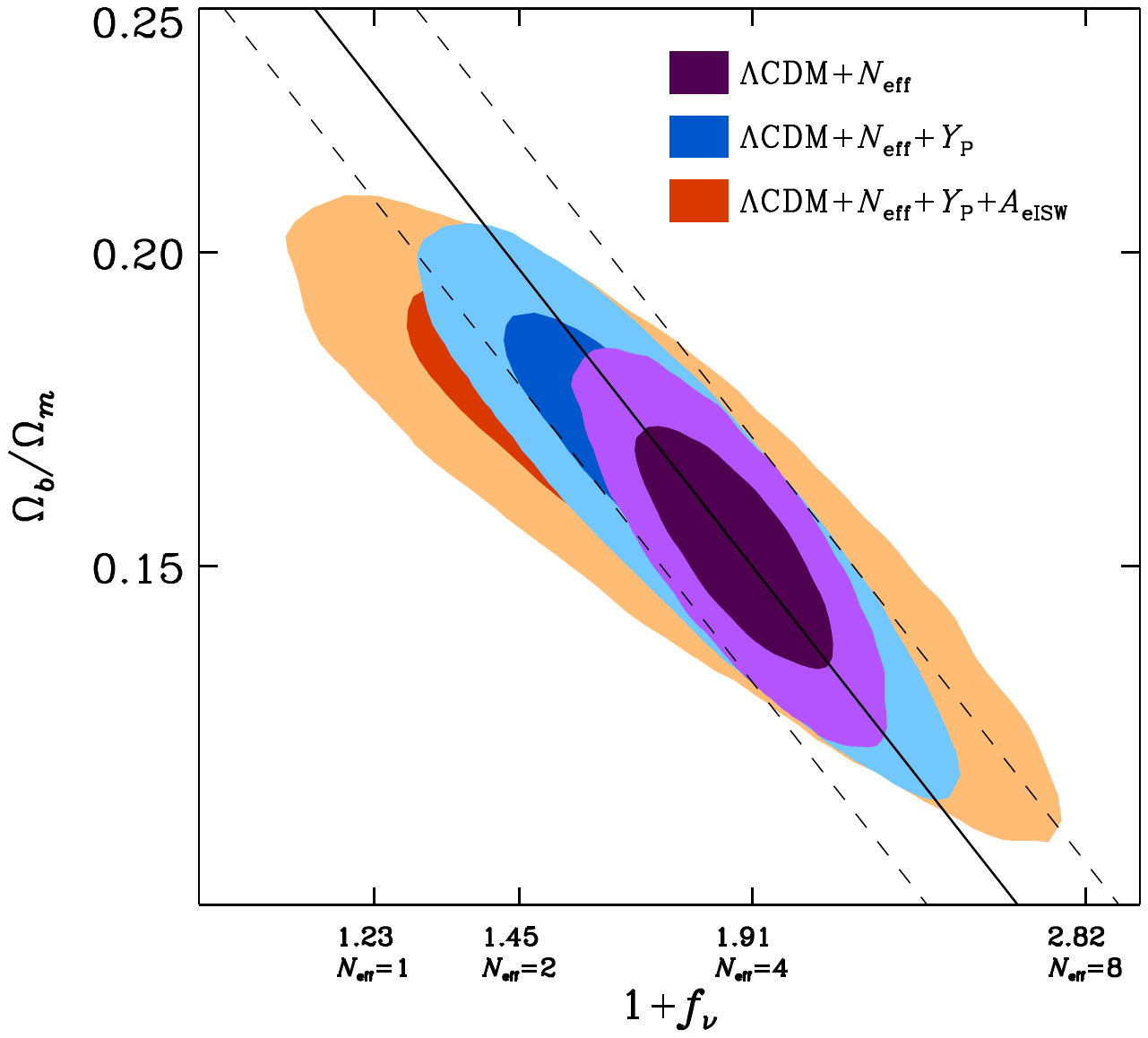}
    \includegraphics[width=0.48\textwidth, trim=2.2cm 13.5cm 5.8cm
        2.8cm]{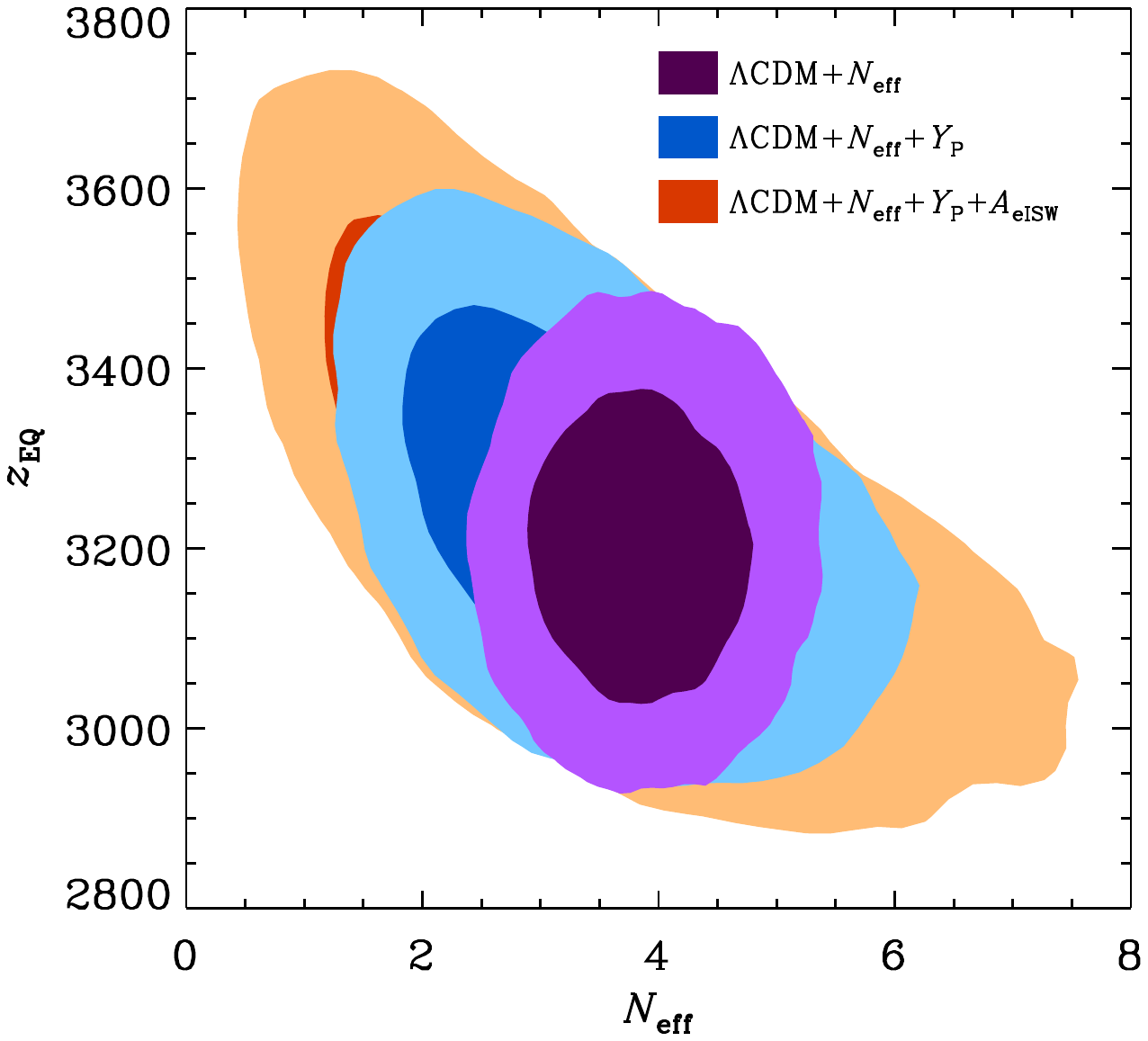}
    \end{center}
    \caption{Probability contours (68\% and 95\%) for the case of BBN consistency
      (purple), free $\yp$ (blue) and free $\yp$ and $A_{\rm eISW}$
      (red). {\em Upper panel:}  baryon fraction, $\Omega_b/\Omega_m$
vs. $(1+f_\nu)$.  
The solid line is for $100\,\Omega_b
h^2=2.267$ and $z_{\rm EQ}=3200$.  Vayring $z_{\rm EQ}$ to 2843 (3557)
we get the right (left)
dashed line. One can see the constant $z_{\rm EQ}$ assumption breaks
down at high baryon fraction.  {\em Lower panel:} the $N_{\rm
  eff}-z_{\rm EQ}$ plane.  Here we can see more directly how the extra
freedom degrades the inerence of $z_{\rm EQ}$.  
        }
    \label{fig:neff_zeq_bf}
\end{figure}

Another effect important for breaking the degeneracy between $\yp$ and $N_{\rm
eff}$ is due
to high baryon fraction.  In our above analysis
we assumed that $z_{\rm EQ}$ is a
fixed constant.  However, this assumption will break down when
$N_{\rm eff}$ decreases to lower values, as it can do if $\yp$ is allowed to
vary freely, while the baryon density remains unchanged.
As $N_{\rm eff}$ decreases, to keep $z_{\rm EQ}$ fixed, $\omega_m$
would decrease, thus driving up the baryon fraction, $\omega_b/\omega_m$.
Effects due to the high baryon fraction make it impossible at
sufficiently low $N_{\rm eff}$ to find a
value of $z_{\rm EQ}$, $\omega_b$ and $\omega_b/\omega_m$ that
reproduce the measured power spectra.  In an attempt to best
accommodate the data, the region of high likelihood departs from 
constant $z_{\rm EQ}$, as can be seen in the upper panel
of Fig.~\ref{fig:neff_zeq_bf} which shows
probability contours in the $(1+f_\nu)$ - $\Omega_b/\Omega_m$ plane.  The lines
are lines of constant $z_{\rm EQ}$ and $\omega_b$.  As extra freedom is
allowed in the model, allowing $N_{\rm eff}$ to extend to lower values, we can
see the expected departure of the probability ridge from the line of constant
$z_{\rm EQ}$ and $\omega_b$, mainly due to variation of $z_{\rm EQ}$.

A high baryon fraction alters the acoustic
dynamics for modes entering the horizon during matter domination,
because a large fraction of the matter sourcing the gravitational
potentials is feeling pressure support.  This pressure support causes
potentials to decay, boosting the amplitude of the fluctuations
as happens with the ``radiation driving'' in the radiation-dominated
era.  At fixed $\omega_b$, increasing $\omega_b/\omega_m$ boosts the
height of the first peak relative to the second.  The existence of this
radiation-driving-like effect in the
matter-dominated era impacts the ability to constrain $z_{\rm EQ}$.
One can see in the lower panel of Fig.~\ref{fig:neff_zeq_bf} how the uncertainty
in $z_{\rm
  EQ}$ increases when $N_{\rm eff}$ is allowed to take on smaller
values (as happens when allowing $\yp$ to vary freely).  The effect is
even greater when allowing further freedom in the model by allowing
$A_{\rm eISW}$ to vary, a parameter we will introduce below.

\section{Constraints on $N_{\rm eff}$} 

Here we present constraints on $N_{\rm eff}$ from current data.  
We begin with our baseline estimates, which are from CMB data alone,
assuming the six-parameter $\Lambda$CDM model extended by one to allow
free $N_{\rm eff}$, with $\yp$ determined by a BBN consistency relation to be
discussed below.  

We also consider departures from our baseline assumptions.  We let
$\yp$ vary away from the standard BBN relationship,
let $dn_s/d\ln k$ vary away from what is expected from
inflation and allow the early ISW effect to be greater
or smaller than expected, by introducing an artificial scaling
parameter, $A_{\rm eISW}$.

Finally, we consider the influence of low-redshift distance
measurements on inferences of $N_{\rm eff}$.

\subsection{Baseline Constraints}

We show our baseline constraints on $N_{\rm eff}$ in the first row of
estimates of Table~\ref{tab:neff_bbn}, which is essentially the same with what
\citet{keisler11} have found for the same model.

\begin{table*}
\begin{center}
\scriptsize
\begin{threeparttable}
\caption{Sensitivity of inferences of $N_{\rm eff}$ and $\omega_b$ to BBN
assumptions}
    \begin{tabular}{c|c|c|c|c|c|c|c|c|c}
    \hline
Assumptions                     & Data                &
$100\,\omega_b^{\rm BBN}$ & $100\,\omega_b^{\rm DEC}$ &
$100\,\Delta \omega_b$ & $N_{\rm eff}^{\rm BBN}$ & $N_{\rm eff}^{\rm
DEC}$ & $\Delta N_{\rm eff}$ & $\omega_m$ &
$10\,\theta_d/\theta_s$\\ \hline

Baseline & CMB & --- & $2.27\pm 0.05$ & --- & --- & $3.86\pm 0.62$ & --- &
$0.148\pm 0.012$ & $1.569\pm 0.015$ \\

${\rm BBN}\neq{\rm DEC}$ & CMB & --- & $2.26\pm 0.05$ & --- & --- &
$3.28^{+1.09}_{-0.89}$ & --- & $0.140^{+0.018}_{-0.015}$ &
$1.569\pm 0.015$ \\

${\rm BBN}\neq{\rm DEC}$ & CMB+$\rm D/H$+$Y^A_{\rm P}$ & $2.29\pm 0.11$ &
$2.26\pm 0.06$ & $-0.023^{+0.123}_{-0.127}$ & $3.80\pm 0.26$ &
$3.76^{+0.77}_{-0.73}$ & $-0.033^{+0.829}_{-0.793}$ & $0.147\pm 0.014$ &
$1.567\pm 0.015$ \\

${\rm BBN}\neq{\rm DEC}$ & CMB+$\rm D/H$+$Y^P_{\rm P}$ & $2.16^{+0.11}_{-0.10}$
& $2.27\pm 0.05$ & $0.105^{+0.118}_{-0.124}$ & $3.09\pm 0.21$ &
$3.93^{+0.79}_{-0.75}$ & $0.843^{+0.829}_{-0.794}$ & $0.149^{+0.015}_{-0.014}$ &
$1.566\pm 0.015$ \\

$\omega_b^{\rm BBN}=\omega_b^{\rm DEC}$ & CMB+$\rm D/H^{PC}$ & --- & $2.27\pm
0.05$ & --- & $3.31\pm 0.58$ & $3.93\pm 0.74$ & $0.617^{+0.896}_{-0.841}$ &
$0.149\pm
0.014$ & $1.567\pm 0.015$ \\

 \hline
    \end{tabular}
    \label{tab:neff_bbn}
    \begin{tablenotes}
	\item The PC superscript indicates the Pettini and Cooke (2012)
inference, A indicates Aver et al. (2011) and P indicates Peimbert et al.
(2007).  The difference $\Delta \omega_b \equiv \omega_b^{\rm DEC} - \omega_b^{\rm BBN}$, and
$\Delta N_{\rm eff} \equiv N_{\rm eff}^{\rm DEC} - N_{\rm eff}^{\rm BBN}$.
    \end{tablenotes}
\end{threeparttable}
\end{center}
\end{table*}

\subsection{Sensitivity to BBN Assumptions}

The standard assumption is that the baryon-to-photon ratio and $N_{\rm
  eff}$ are unchanged from BBN through decoupling.  However, many processes
can change this situation, such as energy injection into the plasma after BBN
which would reduce both $\omega_b$ and $N_{\rm eff}$, or a decay of a massive
dark species into a relativistic dark species after BBN, which would
increase $N_{\rm eff}$ while leaving $\omega_b$ unchanged.  Here we
relax the standard assumption and distinguish the variables with superscripts
DEC (for decoupling) and BBN.

First we consider the extreme of making no assumptions about BBN and
using no light-element abundance data to constrain the BBN
quantities.  In practice this simply means allowing $\yp$ to be free.  As one
would expect from our earlier discussion, allowing
$\yp$ to be a free parameter greatly relaxes the constraints on
$N_{\rm eff}^{\rm DEC}$ because of its impact on $\theta_d/\theta_s$.
This is shown in the second row of Table~\ref{tab:neff_bbn}.  

If we assume standard BBN, we can use measurements of the abundance of
Deuterium relative to Hydrogen, $D/H$, and $\yp$ to determine
$\omega_b^{\rm BBN}$ and $N_{\rm eff}^{\rm BBN}$.  \citet{simha08}
provided fitting formulae for the dependence of $\yp$ and $D/H$ on
these quantities. Here we present 
revised ones \footnote{G. Steigman, private communication}, that incorporate
updates in nuclear reaction rate estimates and a neutron lifetime
estimate.  They are
\bea
\label{eqn:yp}
Y_{\rm P} & = & 0.2381 \pm 0.0006 + \left[\eta_{\rm
    10}+100(S-1)\right]/625 
\eea
and
\bea
\label{eqn:dh}
10^5 D/H &= &2.60(1 \pm 0.06) \left[\frac{6}{\eta_{\rm 10} -
    6(S-1)}\right]^{1.6} 
\eea
where
\bea
\eta_{\rm 10} &\equiv & 10^{10} n_b/n_\gamma = 273.9 \omega_b + 100(S
- 1) \ \ {\rm and} \nonumber \\  
S & =&  \left[1 + 7(N_{\rm eff} - 3.046)/43\right]^{1/2}. 
\eea
To calculate $\yp$ as a function of $\omega_b$ and
 $N_{\rm eff}$ we use the default CosmoMC option which is an
 interpolation over tables produced using PArthENoPE v1.00 described
 in \citet{pisanti08}, only using Eq.~\ref{eqn:yp} for values
outside the bounds of the tables for robustness but very rarely reached.  To
calculate $D/H$ we use Eq.~\ref{eqn:dh}.

For the measurements of light element abundances we mostly follow
\citet{nollett11}.
They assumed, from a compilation of $D/H$ measurements (see \citet{pettini08,
fumagalli11} and references therein)
\be
\log(D/H)  = -4.556 \pm 0.034. 
\ee
We also consider a very recent, and significantly more preicse, $D/H$
measurement by \citet{pettini12} of
\be
\log(D/H)  = -4.596 \pm 0.009.
\ee
Note that signifcant uncertainty arises from nuclear reaction rate
uncertainty.  When using $D/H$ measurements we include the 6\% error
in Eq.~\ref{eqn:dh} and add it in quadrature with the measurement
error.

For Helium \citet{nollett11} considered two different inferences
\bea
Y_P & = & 0.2573 \pm 0.0033 \ \ {\rm (Aver)} \nonumber \\
\yp & = & 0.2477 \pm 0.0029 \ \ {\rm (Peimbert)}
\eea
from \citet{aver11} and \citet{peimbert07} respectively;
we will do the same.

Including the light element abundance measurements to help
constrain $\yp$ reduces the uncertainty in $N_{\rm eff}^{\rm DEC}$,
though it remains comparatively large.  Interestingly, the lower $\yp$
measurement, which is the only one of the two consistent with $N_{\rm
  eff}^{\rm BBN} = 3$, leads to a slightly higher $N_{\rm eff}^{\rm
  DEC}$ inference.  This is because lower $\yp$ values need higher
$N_{\rm eff}$ to get the same $\theta_d/\theta_s$.  

\citet{pettini12} took an inference of the baryon
density from \citet{keisler11}, calculated assuming $N_{\rm eff} = 3.046$, 
combined it with their $D/H$ measurement, and the above
$D/H$ fitting formula, and found $N_{\rm eff}^{\rm BBN} = 3.0 \pm 0.5$.
Thus they show that their data are consistent with the combination of
CMB data and the assumption of $N_{\rm eff}^{\rm BBN} = N_{\rm
  eff}^{\rm DEC} = 3.046$.  

We performed a similar, but different, exercise where we set
$\omega_b^{\rm DEC} = \omega_b^{\rm BBN}$ and estimated both $N_{\rm
  eff}$ values simultaneously from the CMB and D/H data.  There is a
small correlation between $\omega_b$ and $N_{\rm eff}$ as inferred
from the CMB data which leads to increased $\omega_b$ inference when
$N_{\rm eff}$ is allowed to vary.  The net result is that our exercise
leads to a higher value, $N_{\rm eff}^{\rm BBN} = 3.3 \pm 0.6$, with $N_{\rm
eff}^{\rm DEC} = 3.93 \pm 0.74$.

We note that our analysis, in which we allow the BBN quantities to be
different from the DEC quantities, is for a very general scenario and
thefore missing features that may be important for specific scenrios.  For
example, in \citet{eggers12} a scenario was considered in which $N_{\rm
  eff}$ increases after BBN due to the decay of a fraction of the dark
matter into dark radiation.  With such a specific choice, one can
include in the calculation the details of how this conversion
happens over time, and the differences in the dark radiation
perturbations from the case of thermally produced neutrinos.

\subsection{Sensitivity to primordial power spectrum assumptions}

\begin{table*}
\begin{center}
\scriptsize
\begin{threeparttable}
\caption{Sensitivity of inferences of $N_{\rm eff}$ to low-redshift
  distance measurements, $dn_s/d\ln k$ and the
arbitrary amplitude of the early ISW effect}
\label{tab:neff_other}
\begin{tabular}{c|c|c|c|c|c|c|c|c}
\hline
Assumptions                     & Data                &
$100\,\omega_b^{\rm DEC}$ & $N_{\rm eff}^{\rm DEC}$ &
$dn_s/d\ln k$ & $A_{\rm eISW}$ & $\Omega_\Lambda$ & $\omega_m$ &
$10\,\theta_d/\theta_s$\\
\hline
Baseline & CMB & $2.27\pm 0.05$ & $3.86\pm 0.62$ & --- & --- & $0.737\pm 0.025$ &
$0.148\pm 0.012$ & $1.569\pm 0.015$ \\

 & CMB+BAO & $2.25\pm 0.05$ & $3.83\pm 0.60$ & --- & --- & $0.707\pm 0.012$ &
$0.154\pm 0.012$ & $1.569\pm 0.015$ \\

 & CMB+$H_0$ & $2.26\pm 0.04$ & $3.73\pm 0.44$  & --- & ---  & $0.733\pm
0.021$ & $0.147\pm 0.011$ & $1.566\pm 0.012$ \\

 & CMB+BAO+$H_0$ & $2.26\pm 0.04$ & $3.97\pm 0.41$ & --- & --- & $0.708\pm
0.011$ & $0.156^{+0.009}_{-0.008}$ & $1.572\pm 0.011$ \\
\hline

$dn_s/d\ln k$ free & CMB & $2.21\pm 0.07$ & $2.97^{+0.91}_{-0.80}$ &
$-0.025\pm 0.020$ & --- & $0.704\pm 0.041$ & $0.138^{+0.014}_{-0.012}$ &
$1.547^{+0.022}_{-0.021}$ \\

$dn_s/d\ln k$ free & CMB+BAO+$H_0$ & $2.25\pm 0.04$ & $3.76\pm 0.43$ &
$-0.015\pm 0.013$ & --- & $0.706\pm 0.011$ & $0.153\pm 0.009$ & $1.566\pm 0.012$
\\

$A_{\rm eISW}$ free & CMB & $2.29\pm 0.08$ & $3.92\pm 0.65$ & --- &
$0.979\pm 0.055$ & $0.741\pm 0.028$ & $0.149\pm 0.012$ & $1.568\pm 0.015$ \\
\hline
\end{tabular}
\end{threeparttable}
\end{center}
\end{table*}

Another straightforward way to reduce small-scale power is to alter
the primordial power spectrum.  For the usual power-law assumption,
the exponent $n_s$ is sufficiently well determined by low-$\ell$ data that it
cannot mimic the damping effect of $N_{\rm eff}$.
However, if we allow for
$n_s$ to have a logarithmic scale dependence so that $n_s(k) = n_s(k_*) +
\ln(k/k_*)dn_s/d\ln k$ for some constant $dn_s/d\ln k$ then the
resulting power spectra can better mimic the effects of $N_{\rm eff}$.
As a result, if we marginalize over $dn_s/d\ln k$, we increase
the uncertainty in $N_{\rm eff}$, as can be seen in
Table~\ref{tab:neff_other}.   

\subsection{Importance of the early ISW Effect}

\begin{figure}
	\begin{center}
	\includegraphics[width=0.48\textwidth, trim=1.5cm 13.5cm 6.5cm
        2.8cm]{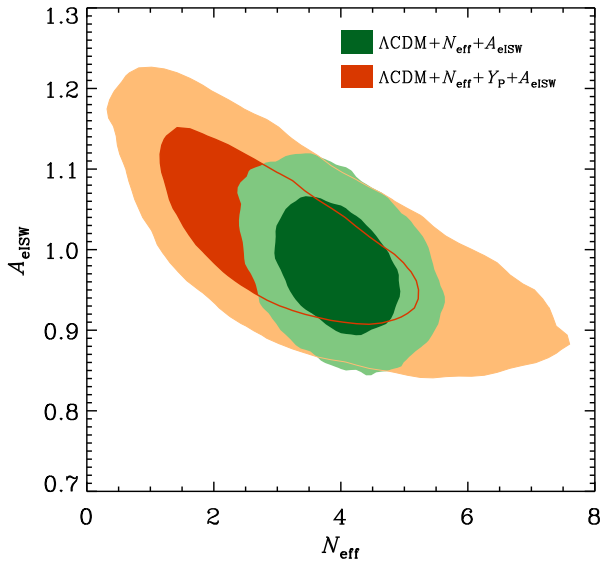}
	\end{center}
	\caption{Probability contours (68\% and 95\%) in the $A_{\rm
            eISW}-N_{\rm eff}$ plane assuming BBN consistency (green, tighter
          contours) and free $\yp$ (red, broader contours).  We see in the
          BBN-consistency case that the constraint on
          $N_{\rm eff}$ persists despite the freedom in the amplitude
          of the eISW effect.  For the case of free $\yp$ the eISW
          effect is important for constraining $N_{\rm eff}$;
          letting the amplitude vary freely degrades the constraints
          on $N_{\rm eff}$ somewhat, especially at low $N_{\rm eff}$.  
        }
	\label{fig:AISW0}
\end{figure}

The anisotropy field today can be written as an integration over
perturbation variables on the past light cone.  Doing so can
be useful both computationally \citep{seljak96} and for analytic understanding.
Writing the Fourier and Legendre-transformed radiation transfer function as such
an integral (ignoring the polarization-dependence of Thomson
scattering for simplicity) one gets 
\bea
\label{eqn:defineAisw}
\Theta_l(k) & =&  \int_0^{\eta_0} d\eta g(\eta)\left[\Theta_0(k,\eta) +
  \Psi(k,\eta)\right] j_l\left[k(\eta_0 - \eta)\right] \nonumber \\
& & -\int_0^{\eta_0} d\eta g(\eta) \frac{i v_b(k,\eta)}{k}
\frac{d}{d\eta}j_l\left[k(\eta_0 - \eta)\right]  \\
& & +\int_0^{\eta_0} d\eta  f(z(\eta),A_{\rm eISW})  e^{-\tau}\left[\dot
  \Psi(k,\eta) - \dot \Phi(k,\eta)\right] \nonumber \\
& & \times  j_l\left[k(\eta_0 -
  \eta)\right] \nonumber.  
\eea
See \citet{dodelsonbook} for definitions.  The final integral is the
so-called integrated Sachs-Wolfe (ISW) effect.  The other terms are
only important when the visibility function, $g(\eta)$ is non-zero,
whereas the ISW term gets contributions along the whole line of sight
between here and recombination where $\tau$ starts to get very large.
The gravitational potential time derivatives are zero for a Universe
dominated by cold dark matter.  They are 
significantly non-zero at early times as the radiation density is still
a significant contributor to the expansion rate (the early ISW effect), and then again at
late times when dark energy becomes important (the late ISW effect). 

In Eq.~\ref{eqn:defineAisw} we have introduced the parameter $A_{\rm
  eISW}$ so we can artificially vary the amplitude of the early ISW
effect.  The function $f(z(\eta), A_{\rm eISW}) = A_{\rm eISW}$ when
the redshift, $z > 30$ and otherwise equals one.  This use of $A_{\rm
  eISW}$ is very similar to the use of $A_{\rm lens}$ to artifically
change the amplitude of the lensing potential power spectrum altering
the CMB power spectrum.  We might have chosen $f$ to go as the square
root of $A_{\rm eISW}$ so that $A_{\rm eISW}$ is scaling the ISW power
(just as $A_{\rm lens}$ scales up the lensing power).  However, a
significant impact of the ISW term in Eq.~\ref{eqn:defineAisw} comes
from its correlation with the other terms in the equation.  Thus there
is no way to make the total contribution of eISW to the power scale as
a single power of $A_{\rm eISW}$.

The greater the amount of radiation around, relative to matter, at
recombination, the greater the amplitude of the early ISW effect.  Thus
increasing $N_{\rm eff}$ (with $\omega_m$ held fixed) would increase
the amplitude of the early ISW effect.  This relationship between $N_{\rm
  eff}$ and the early ISW effect led \citet{bowen02} to cite the ISW effect
as the reason the CMB is sensitive to $N_{\rm eff}$.

To quantitatively investigate the impact of the ISW effect, we could
perform the exercise of turning it off artificially.  But turning off
the ISW effect would so radically change the first peak (dropping it
in power by 28\%) that we instead chose to investigate by letting
$A_{\rm eISW}$ be a free parameter.  If the ISW effect is playing an
important role in constraining $N_{\rm eff}$ then if we let it be a
free parameter, those constraints will degrade and we will also see a
strong correlation between the two parameters.

\begin{figure}
	\begin{center}
	\includegraphics[width=0.48\textwidth, trim=1.5cm 13.5cm 6.5cm
        2.8cm]{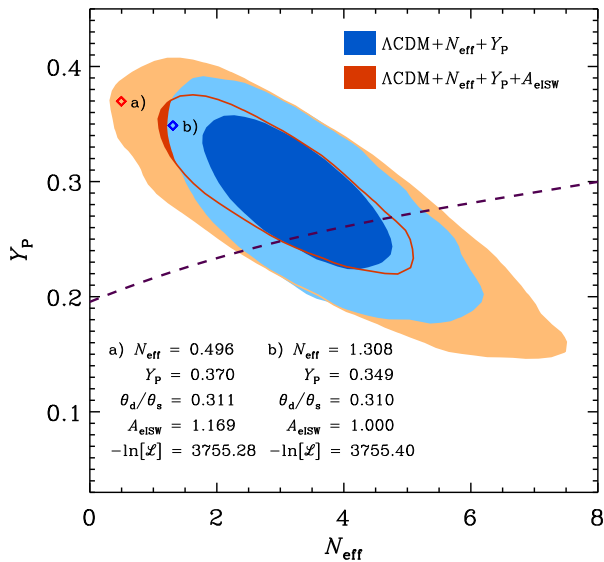}
	\end{center}
	\caption{Allowing $A_{\rm eISW}$ to vary stretches out the
          contours in the $\yp$ plane in the direction not constrained
          by $\theta_d/\theta_s$.  If we assume BBN consistency
          though (the dashed line), then it has very little impact on
          the constraint on $N_{\rm eff}$.  Along the direction of constant
          $\theta_d/\theta_s$, two samples, a) and b), in the two Markov chains
          are picked up with almost identical likelihood values.  The
parameters of interest of these two samples are listed, and the corresponding
power spectra are plotted in Fig.~\ref{fig:AISW2} with the $A_{\rm eISW}$
modulation turned off and on.
        }
	\label{fig:AISW1}
\end{figure}

\begin{figure}
	\begin{center}
	\includegraphics[width=0.48\textwidth, trim=2cm 13.8cm 8.5cm
        8cm]{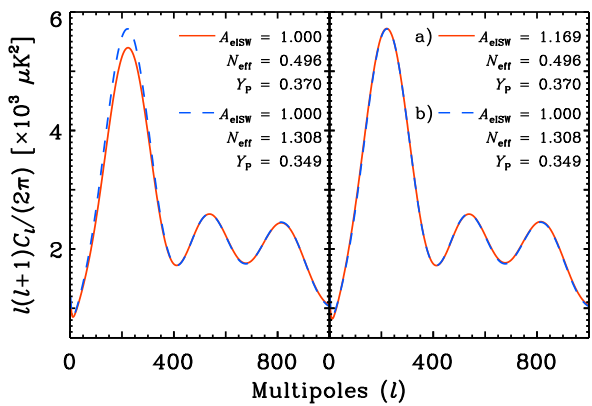}
	\end{center}
	\caption{The power spectra of the two samples shown in
Fig.\ref{fig:AISW1} with the $A_{\rm eISW}$ modulation turned off (on)
in the left
(right) panel.  In this figure we see explicitly how allowing $A_{\rm
            eISW}$ to be artificially free weakens the
          lower bound on $N_{\rm eff}$ in the case that $\yp$ is free.
        }
	\label{fig:AISW2}
\end{figure}

We show probability contours in the $N_{\rm eff}$-$A_{\rm eISW}$ plane
in Fig.~\ref{fig:AISW0} for the case that $\yp$ follows the
BBN-consistency relation and the case that $\yp$ is free.  
For the BBN-consistency case, the constraint on $N_{\rm eff}$
degrades by less than 10\%.  But when $\yp$ is allowed to vary freely,
the $\theta_d/\theta_s$ constraint can no longer determine $N_{\rm eff}$ because
of
the degeneracy with $\yp$.  In this case, the region of the spectrum
affected by ISW is more important for determining $N_{\rm eff}$ so we
see a significant degeneracy between $A_{\rm eISW}$ and $N_{\rm eff}$.

From Fig.~\ref{fig:AISW1} we can see that marginalizing over $A_{\rm
  eISW}$ does indeed loosen up the $N_{\rm eff}$, $\yp$ contour.  One
can see explicitly in Fig.~\ref{fig:AISW2} how letting $A_{\rm eISW}$
vary allows lower $N_{\rm eff}$ and higher $\yp$ than would otherwise
be possible.

We also see from Fig.~\ref{fig:AISW1} that marginalizing over $A_{\rm
  eISW}$ does little to expand the minor axis of the contours.  As we
have emphasized, the constraint along this axis is due to the
constraint on $\theta_d/\theta_s$.  Along the major axis the ISW
effect does play a role in breaking the $N_{\rm eff}-\yp$ degeneracy.
Other effects important in breaking this degeneracy include those
coming from neutrino perturbations \citep{bashinsky04} and high
baryon fraction as discussed in Section~\ref{subsec:constrain_neff_yp}.

%
%

\section{Implications for Low-redshift Distance Measurements}

We now consider the implications of CMB
measurements, in the context of variable $N_{\rm eff}$, for
low-redshift measurements.  

\begin{figure*}
	\begin{center}
	\includegraphics[width=0.98\textwidth, trim=2cm 13.5cm 2cm
        9.0cm]{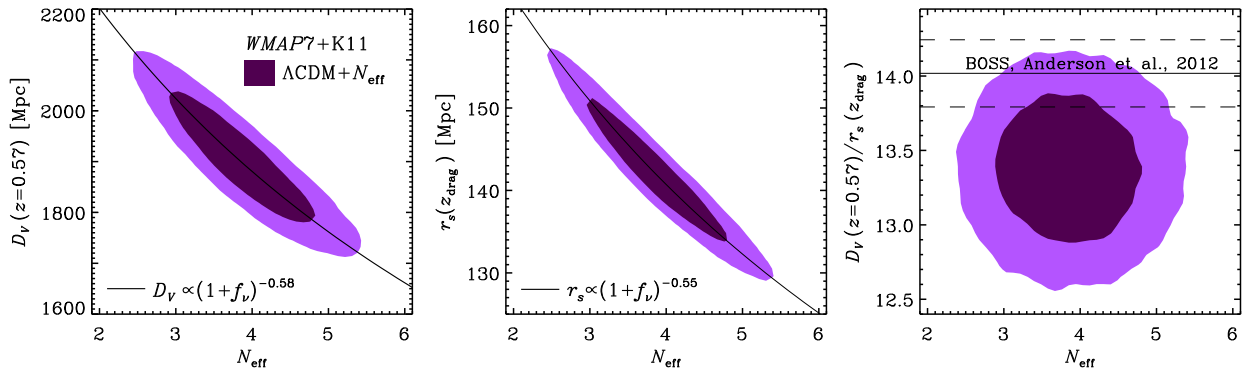}
	\end{center}
	\caption{{\em Left panel:} Contours of constant probability
          indicating the 68\% and 95\% confidence regions in the $N_{\rm
eff}$ - $D_V(z=0.57)$ plane.  The solid line shows the major
correlation direction of $D_V (z=0.57) \propto (1+f_\nu)^{-0.58}$ with the index
best-fitted from the Markov chain.  {\em Middle panel:} Probability
contours in the $N_{\rm eff}$ - $r_s(z_{\rm drag})$ plane; the solid line holds
the best-fit correlation direction $r_s(z_{\rm drag}) \propto
(1+f_\nu)^{-0.55}$.  {\em Right panel:} the relationship between $N_{\rm eff}$
and the ratio, $D_V/r_s$.  The latest BOSS BAO data point \citep{anderson12} is
plotted with the
dashed line showing its $1\sigma$ error, which is re-scaled by
multiplying by the $r_s$-rescaling factor 154.66/150.82.}
	\label{fig:neff_bao}
\end{figure*}

First let's consider the effect on the expansion rate at low
redshift for the $\Lambda$CDM + $N_{\rm eff}$ cosmology.
Increasing $N_{\rm eff}$, with the resulting increase in $\omega_m
\propto (1+f_\nu)$ (to keep $z_{\rm EQ}$ fixed) increases the
expansion rate through the matter-dominated era.  Further, in order to
adjust $D_A$ to keep $\theta_s$ fixed (given the decrease in $r_s$),
the energy density in the cosmological constant, $\omega_\Lambda$,
must increase as well.  With the energy density in both components
important at low redshift increasing, $H(z)$ increases for all
redshifts in the low-redshift (matter-dominated and later) regime.

Note that if we can ignore changes to the limit of integration in
Eq.~\ref{eqn:rs} then we
expect $r_s(z_*) \propto (1+f_\nu)^{-0.5}$.  With this scaling, we then
expect $D_A(z_*) \propto (1+f_\nu)^{-0.5}$ to keep $\theta_s$ fixed.
The only way to achieve that scaling in the $\Lambda$ CDM model 
is to have the dark energy density $\omega_\Lambda$ scale in exactly 
the same way as $\omega_m$.  Hence $H(z)$, and therefore $D_A(z)$
scale the same way at all redshifts.  If $z_{\rm drag}$ also has no
variation with $N_{\rm eff}$ then the net result is that the BAO
observables $D_A(z)/r_s(z_{\rm drag})$ and $H(z)*r_s(z_{\rm drag})$ have 
no dependence on $N_{\rm
  eff}$.  Note that $z_{\rm drag}$ is the epoch of photon-baryon
decoupling defined as in \citet{meiksin99}.  

In practice there are many corrections to the above analysis, e.g.,
the correlation between $z_{\rm drag}$ and $N_{\rm eff}$, but they are
all small and the net result is very little dependence of the BAO
observables on $N_{\rm eff}$.  We can see that in
Table~\ref{tab:neff_other}, where adding in just BAO data makes very
little difference to the $N_{\rm eff}$ inference.  Another consequence
of $\omega_\Lambda$ scaling similarly as $\omega_m$ is that
$\Omega_\Lambda$ does not scale with $N_{\rm eff}$.

Fig.~\ref{fig:neff_bao} quantitatively supports the above
discussion.  We compute the quantity $D_V \equiv
(D_A^2cz/H(z))^{1/3}$ 
at the redshift $z=0.57$
where the latest BAO data were effectively measured \citep{anderson12}.
We see that $D_V (z=0.57) \propto (1+f_\nu)^{-0.58}$, $r_s (z_{\rm
drag}) \propto (1+f_\nu)^{-0.55}$; i.e., they both scale nearly as
$(1+f_\nu)^{-0.5}$.  And their scalings are so similar that the ratio of the two
quantities shows no
noticeable correlation with $N_{\rm eff}$.  In the case of WMAP7+K11+BAO+$H_0$,
we get $\sigma(N_{\rm eff})=0.41$, compared to $\sigma(N_{\rm eff})=0.42$ in K11
\citep{keisler11} for our baseline model.  The only difference comes from the
latest BAO data point which shows a little tension with the $D_V/r_s$ inference
from the CMB data\footnote{Note that \citet{anderson12} report
  $D_V/r_s$ for an $r_s$ given by the fitting formula of
  \citet{eisenstein98}.  We have adjusted the reported value for
  comparison to $D_V/r_s$ calculated with our definition of $r_s$ by
  multiplying it by 154.66/150.82.  See \cite{mehta12}}.

What {\em is} sensitive to $N_{\rm eff}$ at low redshift are absolute
distance measures that are not calibrated with $r_s$.  
The inference of $H_0$, given $\Lambda$CDM (with standard radiation content)
calibrated with the WMAP7 and SPT data is $h = 0.710 \pm 0.021$.  This is
1.17-$\sigma$ lower than the Riess et al. 2011 value of $h = 0.738 \pm
0.024$ \citep{riess11}.  Increasing $N_{\rm eff}$ to the higher value
preferred by the CMB data alone brings these two
inferences into better agreement.  One can see the impact of the Riess
et al. measurement on the reduced uncertainty in $N_{\rm eff}$ in the
Table entries for the $\Lambda$CDM + $N_{\rm eff}$ cosmology;
including the Riess et al. measurement drops the uncertainty by about 50\%.

Varying $N_{\rm eff}$ thus changes the $\Lambda$CDM, CMB-calibrated
predictions for $H_0$, without much change in the predictions
for the BAO data -- consistent with the analysis in
\citet{eisenstein04}.  
Increasing $N_{\rm eff}$ thus can eliminate the
tension between the BAO data and the Riess et al. $H_0$ measurement as
pointed out in \citet{mehta12}.  

The case of $dn_s/d\ln k$ free allows us to see this preference of the
low-redshift data for the $H(z)$ that comes from an
increased $N_{\rm eff}$.  With CMB data alone, letting the running
vary leads to a downward shift in the central value for $N_{\rm eff}$.
Adding in the $H_0$ and BAO measurements shifts the preferred value
for $N_{\rm eff}$ back up towards 4.

Increasing $N_{\rm eff}$ also has implications for the growth of
structure.  Increasing $N_{\rm eff}$ leads to an increase in
$\omega_m$ which in turn decreases $\omega_b/\omega_m$.  Decreasing
the baryon fraction decreases the pressure support felt by matter
prior to recombination, thereby boosting the growth of structure on
scales smaller than the sound horizon at recombination; i.e., scales
smaller than about 150 Mpc \citep{eisenstein98}.  Therefore increasing $N_{\rm
eff}$
increases cluster abundances, that are sensitive to the power spectrum
amplitude on $\sim$ 10 Mpc scales.  This effect can be seen in joint 
estimates of $m_\nu$ (which has the opposite impact on small-scale
power) and $N_{\rm eff}$ from cluster abundances in
\citet{benson11} and also in \cite{hamann11}.

\section{Conclusion}

There are several ways that massless neutrinos impact the anisotropy of the
cosmic microwave background.  Here we have shown that current CMB
constraints on $N_{\rm eff}$ are dominated by the impact of the
neutrino energy density on the expansion rate.  Although $\theta_s$ is
sensitive to this change in expansion rate from $N_{\rm eff}$, its simultaneous
sensitivity to the distance to last scattering greatly limits how well
$N_{\rm eff}$ can be reconstructed from it alone.  Measuring the damping
tail region has allowed a measurement of $\theta_d$ as well.  Since
the response of $\theta_d$  to the expansion rate is different from
the response of $\theta_s$, their ratio (which is independent of the
distance to last scattering) is sensitive to the expansion rate. 

The above analysis assumes that $\yp$ follows the BBN consistency
relation.  Since $\yp$ also alters $\theta_d$, allowing it to vary
freely introduces a near degeneracy between $N_{\rm eff}$ and $\yp$.
This near degeneracy is broken by a number of physical effects at low
$N_{\rm eff}$, including the early ISW effect and effects of a high baryon fraction.  
At high $N_{\rm eff}$ the acoustic oscillation phase shifts probably
play an important role, although we have not quantitatively confirmed
this hypothesis.

We defined a phenomenological scaling parameter of the early ISW effect, $A_{\rm
eISW}$.  We used it to show that the ISW effect plays a very small
role for the case of BBN consistency, although, as just noted above, it
does contribute to constraints on $N_{\rm eff}$ when $\yp$ is allowed
to vary freely.  We found that, assuming BBN consistency, $A_{\rm
  eISW} = 0.979\pm 0.055$ -- a highly significant (though model
dependent) detection of the early ISW effect.

We tested the consistency of inferences of $\omega_b$, $N_{\rm eff}$ and $\yp$
from CMB data, with inferences from light element abundance
measurements and BBN theory.  We see no strong evidence for any
inconsistency, though the bounds are quite loose.  For the most
discrepant case we find $N_{\rm eff}^{\rm DEC} - N_{\rm eff}^{\rm BBN}
= 0.84^{+0.83}_{-0.79}$.  We note that a
simultaneous inference of $N_{\rm eff}^{BBN}$ and $N_{\rm eff}^{\rm DEC}$ from
the CMB and the new Pettini \& Cooke D/H measurement results in an $N_{\rm eff}^{\rm
BBN}$ more consistent with 4 than is the case for the similar exercise
performed in \citet{pettini12}.

We considered the impact of low-redshift distance-redshift relation
measurements on determination of $N_{\rm eff}$.  We found that BAO
data are not very sensitive to $N_{\rm eff}$ because $D_V$ and $r_s$
scale similarly with $N_{\rm eff}$.  The same cancellation does not
occur for distance measures that calibrate independent of the CMB,
such as the Hubble constant determination in \citet{riess11}.  

We will have tighter measurements from the forthcoming analysis of the
entire SPT survey, and we expect improvements to come from
{\it Planck} in early 2013. We project that the error on $N_{\rm{eff}}$ will
reduce to $\sim$ 0.33 using simulated full-survey SPT data combined
with existing WMAP and $H_0$ data, and will reduce further to
$\sim$ 0.20 using simulated {\it Planck} data and existing $H_0$ data, assuming
the {\it Planck} data and foreground model of \citet{millea11}, consistent with
the forecasts of \citet{bashinsky04} and \citet{hannestad06}.

\begin{acknowledgments}
  We thank A. Albrecht, B. Benson, O. Dor\'{e}, E. Komatsu, M. Luty,
  J. Ruhl, L. Strigari, A. Vikhlinin and M. White for useful
  conversations.  We acknowledge support from NSF awards no. 0709498
  and ANT-0638937.  We used CosmoMC \citep{lewis02}.
\end{acknowledgments}

\bibliography{Neff}

\end{document}